\documentclass[letterpaper,english,aps,prl,reprint,groupedaddress,showpacs]{revtex4-1}
\usepackage[T1]{fontenc}
\usepackage[latin9]{inputenc}
\usepackage{amsmath}
\usepackage{amssymb}
\usepackage{graphicx}

\makeatletter

\pdfpageheight\paperheight
\pdfpagewidth\paperwidth

\usepackage[T1]{fontenc} \usepackage{mathptmx}

\@ifundefined{showcaptionsetup}{}{%
 \PassOptionsToPackage{caption=false}{subfig}}
\usepackage{subfig}
\makeatother

\usepackage{babel}
\begin{document}
\captionsetup[figure]{name={FIG.},labelsep=period}

\preprint{APS/123-QED}

\title{Trailing Waves}

\author{Yongsheng Shao}

\author{Liang Zeng}

\affiliation{Shaanxi Key Laboratory of Mechanical Product Quality Assurance and
Diagnostics,~\\
Xi\textquoteright an, Shaanxi Province, China, 710049 }

\collaboration{School of Mechanical Engineering, Xi'an Jiaotong University}

\author{Jing Lin}
\email{Corresponding author:jinglin@mail.xjtu.edu.cn}

\affiliation{State Key Laboratory for Manufacturing Systems Engineering Xi\textquoteright an
Jiaotong University,~\\
Xi\textquoteright an, Shaanxi Province, China, 710049 }

\collaboration{School of Mechanical Engineering, Xi'an Jiaotong University}

\date{\today}
\begin{abstract}
We report a special phenomenon: trailing waves. They are generated
by the propagation of elastic waves in plates at large frequency-thickness
($fd$) product. Unlike lamb waves and bulk waves, trailing waves
are a list of non-dispersive pulses with constant time delay between
each other. Based on Raleigh-Lamb equation, we give the analytical
solution of trailing waves under a simple assumption. The analytical
solution explains the formation of not only trailing waves but also
bulk waves. It helps us better understand elastic waves in plates.
We finally discuss the great potential of trailing waves for nondestructive
testing.
\end{abstract}



\maketitle

\begin{figure}[b]
\subfloat[\label{fig:1a}]{\includegraphics[scale=0.5]{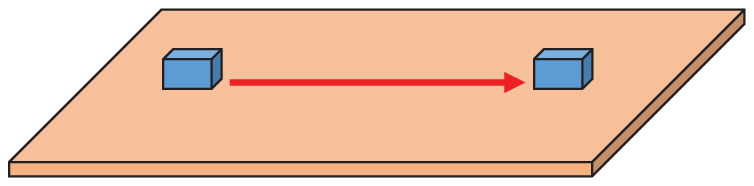}}\subfloat[\label{fig:1b}]{\includegraphics[scale=0.5]{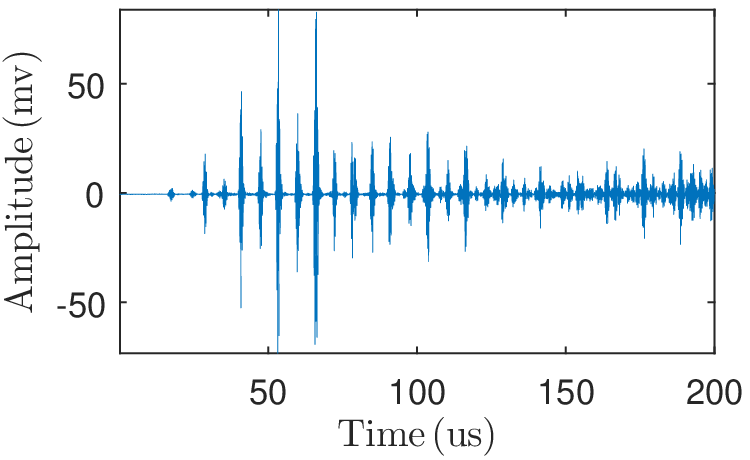}

}\caption{\label{fig:1} (a) Experimental setup. Two ultrasonic transducers
are placed on the surface of the plate. They are employed as the actuator
and receiver respectively. (b) The received trailing waves. The time
delay between pulses remains unchanged.}
\end{figure}

Bulk waves and Lamb waves are widely used in non-destructive testing
for block structures and thin plates respectively\citep{RN244,RN3403,RN341,RN3471,RN3596,RN3603,RN3595}.
Lamb waves are generated at small $fd$ product while bulk waves are
generated at extremely large $fd$ product. However, the transition
state between bulk and Lamb waves has rarely been studied. We have
observed a special phenomenon during the detection of thick plates
and called it trailing waves. Trailing waves are believed to be the
transition state between bulk and Lamb waves. As shown in FIG.\ref{fig:1a},
a 40-mm-thick aluminum plate was excited by an ultrasonic transducer
with a toneburst signal with the center frequency of 3 MHz. The received
signal (see FIG.\ref{fig:1b}) is a list of stable longitudinal pulses
with constant time delay and travels at the velocity of longitudinal
waves. 
 Actually, a similar phenomenon is previously reported and generated
by edge excitation\citep{RN3163,RN3167,RN3400,RN3479,RN1462}, which
is treated as a special case of trailing waves in this paper later. 

Consider elastic waves propagating in a free plate, on the upper and
lower surface of which the traction force vanishes. Equations of this
problem is known as Raleigh-Lamb frequency relations, which can be
expressed as follows: 

\begin{equation}
\frac{\tan qh}{\tan ph}=-\frac{4k^{2}pq}{(q^{2}-k^{2})^{2}}\label{eq:1}
\end{equation}

\begin{equation}
\frac{\tan qh}{\tan ph}=-\frac{(q^{2}-k^{2})^{2}}{4k^{2}pq}\label{eq:2}
\end{equation}
Where $h$ is half of the thickness $d$. $p$ and $q$ are given
by

\begin{equation}
q^{2}=\frac{\omega^{2}}{c_{T}^{2}}-k^{2}\label{eq:3}
\end{equation}

\begin{equation}
p^{2}=\frac{\omega^{2}}{c_{L}^{2}}-k^{2}\label{eq:4}
\end{equation}
Here $c_{T}$ is the velocity of transverse waves, $c_{L}$ is the
velocity of longitudinal waves and wavenumber $k$ is equal to $\omega/c_{p}$,
where $c_{p}$ is the phase velocity and $\omega$ is the circular
frequency. Eqs. \eqref{eq:1} and \eqref{eq:2} are transcendental
equations, which govern the symmetric and antisymmetric modes respectively.
These equations can determine the velocities that waves propagate
at within plates under a particular $fd$ product. Traditionally,
they can be solved only by numerical methods, simple though they are\citep{Rose2014Ultrasonic}.
We believe that trailing waves are the solutions of Raleigh-Lamb equation
at large $fd$ product. Therefore, we focus on the solutions under
large $fd$ product and get the approximate analytical solutions under
a simple assumption. Firstly, we consider the symmetric modes. In
the dispersion curve, the nearly non-dispersive region where $c_{p}\thickapprox c_{L}$
is considered and the region gets larger by the increasing of $fd$
product. Based on this consideration, the value of $k/\left(\omega/c_{L}\right)$
is very close to unity and Eq. \eqref{eq:3} becomes 
\begin{equation}
q^{2}=\frac{\omega^{2}}{c_{L}^{2}}(\frac{c_{L}^{2}}{c_{T}^{2}}-1)\label{eq:5}
\end{equation}
Substituting Eq. \eqref{eq:5} into Eq. \eqref{eq:1}, we can write
\begin{equation}
ph\tan ph=-\frac{h}{\varepsilon}\gamma\label{eq:6}
\end{equation}
Where 

\begin{equation}
\varepsilon=4\frac{c_{d}}{\omega}\frac{\sqrt{\left(c_{L}/c_{t}\right)^{2}-1}}{\left[\left(c_{L}/c_{t}\right)^{2}-2\right]^{2}}=\frac{c_{L}}{\omega}\frac{\left(1-2\sigma\right)^{\frac{3}{2}}}{\sigma^{2}},\gamma=\tan qh\label{eq:7}
\end{equation}
Here $\sigma$ is Poisson's ratio. Considering large $fd$ product,
we take the assumption that $\frac{h}{\varepsilon}\gamma$ is very
large. This assumption will be inaccurate only in a few cases where
$\gamma$ is small. This limitation weakens by the increase of $fd$
product ($\frac{h}{\varepsilon}$ gets larger).  Let assume that
the form of solution is $ph=r+is+\left(n+\frac{1}{2}\right)\pi$ where
$r$ and $s$ are small. We get that$\tan\left(ph\right)\thickapprox-\left(r+is\right)^{-1}$,
which accords with the right hand side of Eq.\ref{eq:6}. The term
$n\pi$ meets the periodicity of tangent function. Eq. \eqref{eq:6}
may be written as
\begin{equation}
r+is+\left(n+\frac{1}{2}\right)\pi\thickapprox\frac{h\gamma}{\varepsilon}\left(r+is\right)
\end{equation}
Which gives 
\begin{eqnarray}
r\thickapprox\frac{\left(n+\frac{1}{2}\right)\pi}{\gamma h/\varepsilon-1}\thickapprox\frac{\varepsilon}{h\gamma}\left(n+\frac{1}{2}\right)\pi & , & s\thickapprox0
\end{eqnarray}
According to Eq. \eqref{eq:4}, the phase velocity of symmetric modes
is 
\begin{equation}
c_{p,s}=\left[\frac{1}{c_{L^{2}}}-\left(n+\frac{1}{2}\right)^{2}\pi^{2}\left(h\omega-\frac{c_{L}}{\gamma}\frac{\left(1-2\sigma\right)^{\frac{3}{2}}}{\sigma^{2}}\right)^{-2}\right]^{-\frac{1}{2}}\label{eq:10}
\end{equation}
where $n$ represents different modes of trailing waves. The group
velocity $c_{g}$ can be obtained by the relation $c_{g}=c_{p}^{2}\left(c_{p}-\omega\frac{\mathrm{d}c_{p}}{\mathrm{d}\omega}\right)^{-1}$.
The solution of anti-symmetric modes is almost the same as above.
Therefore, the phase velocity of anti-symmetric modes is
\begin{equation}
c_{p,a}=\left[\frac{1}{c_{L^{2}}}-n^{2}\pi^{2}\left(h\omega+\gamma c_{L}\frac{\left(1-2\sigma\right)^{\frac{3}{2}}}{\sigma^{2}}\right)^{-2}\right]^{-\frac{1}{2}}\label{eq:11}
\end{equation}
 
\begin{figure}[b]
\subfloat[\label{fig:phase-velocity}]{\includegraphics{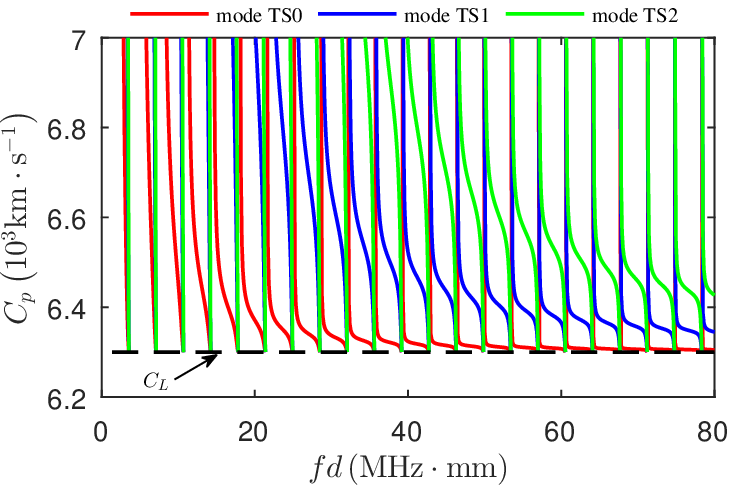}}\\
\subfloat[\label{fig:group-velocity}]{\includegraphics{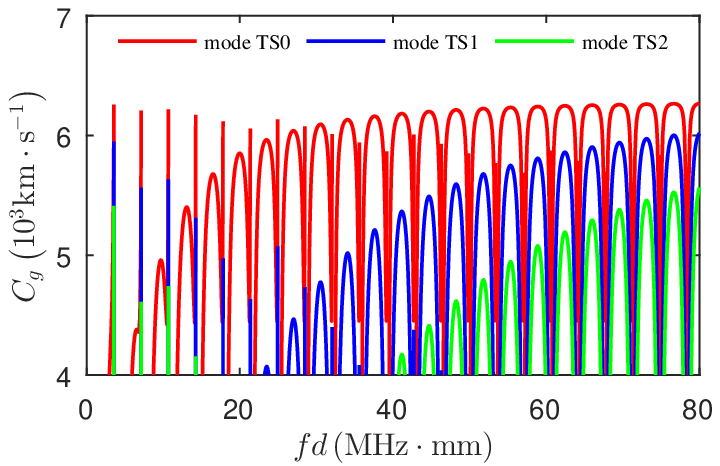}}\caption{\label{fig:epsart-1} Dispersion curves of trailing waves (symmetric
modes). (a) The phase velocity. (b) The group velocity.}
\end{figure}
\begin{figure}[b]
\subfloat[]{\includegraphics[scale=0.5]{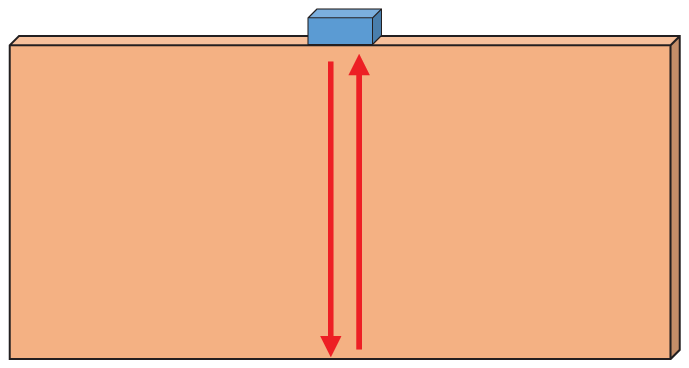}

}\subfloat[]{\includegraphics[scale=0.5]{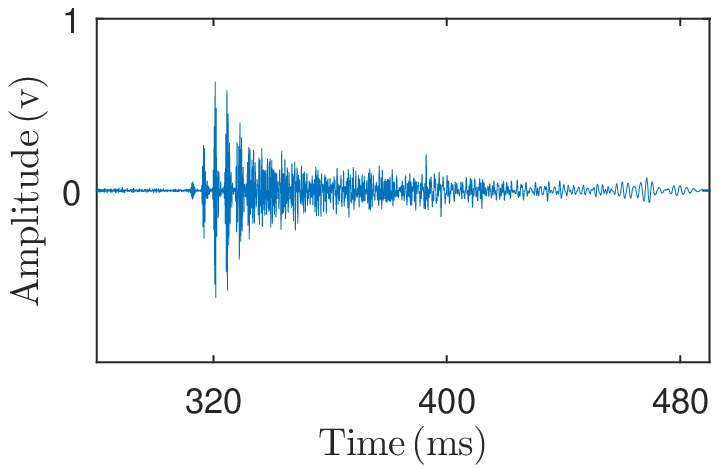}}\\
\subfloat[\label{fig:3c}]{\includegraphics{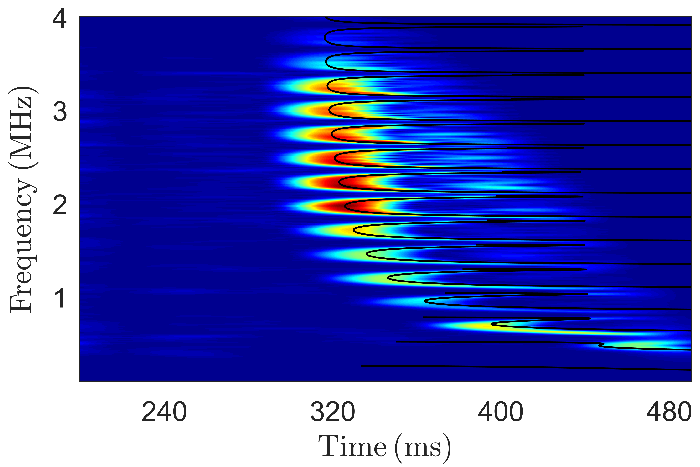}}\caption{\label{fig:epsart-1-1} (a) Experimental setup. One ultrasonic transducer
is employed as the Self-Transmitting and Self-Receiving sensor. The
sensor is excited by a broadband toneburst signal with the frequency
from 25 KHz to 4 MHz. (b) The received trailing waves of symmetric
modes. (b) The time-frequency spectrogram of the experimental signal,
compared with TS0 mode (black line).}
\end{figure}

FIG. \ref{fig:phase-velocity} and FIG. \ref{fig:group-velocity}
show the value of $c_{p}$ and $c_{g}$ of symmetric modes as a function
of $fd$ product. The singular value of group velocity is due to numerical
instability. At the region where $c_{p}\thickapprox c_{L}$, the group
velocity arrives its local maximum. While at the region where the
phase velocity changes rapidly, the group velocity approaches zero
and the corresponding frequencies are called the cut-off frequencies
$f_{c}$. Note that the definition of trailing wave modes is quite
different from the Lamb wave modes. Classical Lamb wave modes are
separated by $f_{c}$ while trailing wave modes are determined by
variable $n$ in Eqs. \eqref{eq:10} and \eqref{eq:11}. Each trailing
wave mode consists of a cluster of adjacent Lamb wave modes. These
differences are then demonstrated by an experiment. In our experiment,
the symmetric modes of trailing waves were obtained by edge excitation
(FIG. \ref{fig:epsart-1-1}). Geometric symmetry excitation could
eliminate the antisymmetric modes. This means the similar phenomenon
previously reported as secondary signals etc. is actually the symmetric
modes of trailing waves. In the time-frequency spectrogram (\ref{fig:3c}),
the experimental trailing waves are consistent with the theoretical
result (TS0 mode) while higher modes are obscure. 

\begin{figure}
\includegraphics{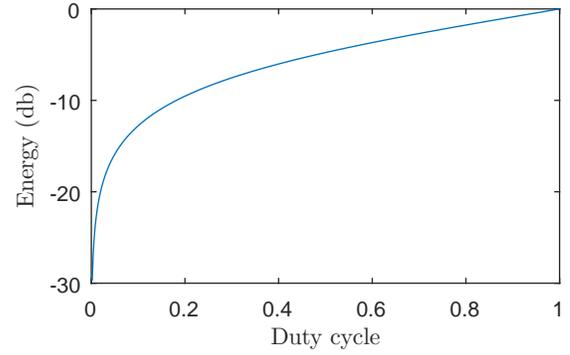}\caption{\label{fig:The-energy-of}The energy of first pulse.}

\end{figure}

Next, the formation of trailing waves will be discussed based on the
analytical solution. We will mainly analyze the symmetric modes. As
expressed in Eq. \eqref{eq:10}, the modulation factor $h\omega$
(MAF) and the periodic factor $\gamma$ (PAF) play key roles in the
formation of trailing waves. Expanding Eq. \eqref{eq:7}, we get
\begin{equation}
\gamma=\tan qh=\tan\left(2\pi fh\sqrt{\frac{1}{c_{T}^{2}}-\frac{1}{c_{p}^{2}}}\right)\label{eq:12}
\end{equation}
with the period of $\triangle f=1/\left(2h\sqrt{\frac{1}{c_{T}^{2}}-\frac{1}{c_{p}^{2}}}\right)$.
PAF forces the phase and group velocity repeating in the period of
$\triangle f$ in frequency domain. According to Fourier transform,
the periodicity in frequency domain leads to uniformly-spaced sampling
in time domain, which means the waves will split into a list of waves
with constant time delay. The time delay, determined by PAF, is $\Delta t=1/\Delta f$.
Additionally, the cut-off frequency $f_{c}$ is $f_{c}=n/\left(2h\sqrt{\frac{1}{c_{T}^{2}}-\frac{1}{c_{p}^{2}}}\right)$.
The group velocity approaches 0 near $f_{c}$ because the phase velocity
goes infinite. On the other hand, the amplitude distribution of trailing
waves is mainly determined by MAF. As the $fd$ ($h\omega$) product
goes larger, the region where $c_{p}\thickapprox c_{L}$ becomes larger,
which makes the result more accurate; meanwhile, a larger range of
group velocity becomes closer to $c_{L}$. The problem of amplitude
distribution can be regarded as a reverse process of Fourier transform
of periodic signal that repeats in the time domain. We simplified
the group velocity dispersion curve of trailing waves into periodic
rectangular pulses $F\left(\omega\right)$ that repeat in the frequency
domain. The period of $F\left(\omega\right)$ is determined by $\triangle f$
: $\omega_{T}=2\pi\triangle f$. The duty cycle ($\omega_{t}/\omega_{T}$)
is qualitatively described by and positively associated with the region
where $c_{g}$ is close to $c_{L}$. The amplitude distribution of
trailing waves can be obtained by inverse Fourier transform of $F\left(\omega\right)$:
\begin{align}
\ensuremath{f_{n}} & =\frac{1}{\omega_{T}}\int_{-\frac{\omega_{T}}{2}}^{\frac{\omega_{T}}{2}}F\left(\omega\right)e^{jn\omega t_{0}}d\omega\nonumber \\
 & =\frac{\omega_{t}}{\omega_{T}}{\rm Sa}\left(\frac{nt_{0}\omega_{t}}{2}\right),n=0,\pm1,\pm2\ldots
\end{align}
Where $t_{0}=\frac{2\pi}{\omega_{T}}$ and $\ensuremath{{\rm Sa}}$
is sampling function. It is noted that although the negative part
of time delay $nt_{0}$ has no physical meaning, the actual amplitude
distribution of the pulses in the trailing waves equals to the sum
of both negative and positive part, which is

\begin{equation}
f\left(n\right)=\begin{cases}
f_{n} & n=0\\
f_{n}+f_{-n} & n=1,2,3\ldots
\end{cases}\label{eq:14}
\end{equation}
Actually, Eq. \eqref{eq:14} can not quantitatively described the
amplitude distribution since the duty cycle is a qualitative description
variable for $c_{g}$. However, qualitative equation still has the
ability to analyze the amplitude distribution of trailing waves. From
Eq. \eqref{eq:14}, it can be seen that the amplitude of first pulse
increases with the duty cycle. As shown in FIG. \ref{fig:The-energy-of},
the first pulse contains the most energy of trailing waves when duty
cycle approaches one, which happens at large $fd$ product. 
Under this condition, the secondary pulses of trailing waves vanish
and the trailing waves are simplified to longitudinal bulk waves.
This indicates that Lamb waves, trailing waves and bulk waves are
three kind of waves in traction-free plates that governed by Raleigh-Lamb
equation.  
\begin{figure}
\includegraphics{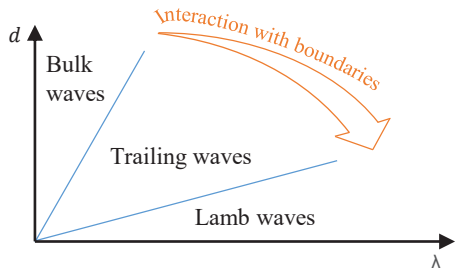}\caption{\label{fig:interaction}Forming conditions of bulk waves, trailing
waves and Lamb waves. The interaction with boundaries will be enhanced
by the increasing of $\lambda/d$.}
\end{figure}
The major formative factor of these three kind of waves is the ratio
of wavelength to thickness ($\lambda/d$), which determines how waves
interact with boundaries (FIG.\ref{fig:interaction}). The incident
waves in infinite media, where boundaries have no influence on wave
propagation, generate bulk waves. Trailing waves are generated when
incident waves interact several times with boundaries. The waves
reflecting and refracting with boundaries for infinite times generate
Lamb waves. 

Besides Lamb waves and bulk waves, trailing waves provide a completely
new way for ultrasonic-based damage detection, with unique advantages
for thick plates. Generally, the detection of thick plates can be
completed by surface scan of bulk waves. High resolution can be achieved
by bulk waves, but the detection efficiency will be low because only
a small area can be detected by single excitation. On the other hand,
there will be limitations if choosing Lamb waves. Although the detection
by Lamb waves can be completed rapidly, it brings great difficulties
under large $fd$ product. High frequency results in high resolution
while more complex and more modes will be generated; low frequency
leads to much worse resolution than bulk waves. With high excitation
frequency and guided-wave-like propagation mode, trailing waves are
able to achieve high resolution and high detection efficiency at the
same time. Meanwhile, the disadvantage of trailing-waves-based damage
detection is: a pulse train of reflection rather than a single pulse
will complicate the interpretation of signals. 

In summary, we obtain the analytical solution of trailing waves by
Raleigh-Lamb equation at large $fd$ product. Based on PAF and MAF,
we explain the formation of trailing waves and bulk waves. Compared
with the symmetric modes of trailing waves that generated in the experiment,
our result is validated. Based on the analysis, we find that not only
Lamb waves but also trailing waves and bulk waves are governed by
Raleigh-Lamb equation. Trailing waves also have great potential on
resolution and detection efficiency for damage detection of thick
plates.

\bibliographystyle{apsrev4-1}
\bibliography{trailingwaves}

\begin{thebibliography}{13}%
\makeatletter
\providecommand \@ifxundefined [1]{%
 \@ifx{#1\undefined}
}%
\providecommand \@ifnum [1]{%
 \ifnum #1\expandafter \@firstoftwo
 \else \expandafter \@secondoftwo
 \fi
}%
\providecommand \@ifx [1]{%
 \ifx #1\expandafter \@firstoftwo
 \else \expandafter \@secondoftwo
 \fi
}%
\providecommand \natexlab [1]{#1}%
\providecommand \enquote  [1]{``#1''}%
\providecommand \bibnamefont  [1]{#1}%
\providecommand \bibfnamefont [1]{#1}%
\providecommand \citenamefont [1]{#1}%
\providecommand \href@noop [0]{\@secondoftwo}%
\providecommand \href [0]{\begingroup \@sanitize@url \@href}%
\providecommand \@href[1]{\@@startlink{#1}\@@href}%
\providecommand \@@href[1]{\endgroup#1\@@endlink}%
\providecommand \@sanitize@url [0]{\catcode `\\12\catcode `\$12\catcode
  `\&12\catcode `\#12\catcode `\^12\catcode `\_12\catcode `\%12\relax}%
\providecommand \@@startlink[1]{}%
\providecommand \@@endlink[0]{}%
\providecommand \url  [0]{\begingroup\@sanitize@url \@url }%
\providecommand \@url [1]{\endgroup\@href {#1}{\urlprefix }}%
\providecommand \urlprefix  [0]{URL }%
\providecommand \Eprint [0]{\href }%
\providecommand \doibase [0]{http://dx.doi.org/}%
\providecommand \selectlanguage [0]{\@gobble}%
\providecommand \bibinfo  [0]{\@secondoftwo}%
\providecommand \bibfield  [0]{\@secondoftwo}%
\providecommand \translation [1]{[#1]}%
\providecommand \BibitemOpen [0]{}%
\providecommand \bibitemStop [0]{}%
\providecommand \bibitemNoStop [0]{.\EOS\space}%
\providecommand \EOS [0]{\spacefactor3000\relax}%
\providecommand \BibitemShut  [1]{\csname bibitem#1\endcsname}%
\let\auto@bib@innerbib\@empty
\bibitem [{\citenamefont {Li}\ \emph {et~al.}(2013)\citenamefont {Li},
  \citenamefont {Pain}, \citenamefont {Wilcox},\ and\ \citenamefont
  {Drinkwater}}]{RN244}%
  \BibitemOpen
  \bibfield  {author} {\bibinfo {author} {\bibfnamefont {C.}~\bibnamefont
  {Li}}, \bibinfo {author} {\bibfnamefont {D.}~\bibnamefont {Pain}}, \bibinfo
  {author} {\bibfnamefont {P.~D.}\ \bibnamefont {Wilcox}}, \ and\ \bibinfo
  {author} {\bibfnamefont {B.~W.}\ \bibnamefont {Drinkwater}},\ }\href
  {\doibase DOI 10.1016/j.ndteint.2012.07.006} {\bibfield  {journal} {\bibinfo
  {journal} {Ndt I\& E International}\ }\textbf {\bibinfo {volume} {53}},\
  \bibinfo {pages} {8} (\bibinfo {year} {2013})}\BibitemShut {NoStop}%
\bibitem [{\citenamefont {Su}\ \emph {et~al.}(2006)\citenamefont {Su},
  \citenamefont {Ye},\ and\ \citenamefont {Lu}}]{RN3403}%
  \BibitemOpen
  \bibfield  {author} {\bibinfo {author} {\bibfnamefont {Z.~Q.}\ \bibnamefont
  {Su}}, \bibinfo {author} {\bibfnamefont {L.}~\bibnamefont {Ye}}, \ and\
  \bibinfo {author} {\bibfnamefont {Y.}~\bibnamefont {Lu}},\ }\href {\doibase
  10.1016/j.jsv.2006.01.020} {\bibfield  {journal} {\bibinfo  {journal}
  {Journal of Sound and Vibration}\ }\textbf {\bibinfo {volume} {295}},\
  \bibinfo {pages} {753} (\bibinfo {year} {2006})}\BibitemShut {NoStop}%
\bibitem [{\citenamefont {Drinkwater}\ and\ \citenamefont
  {Wilcox}(2006)}]{RN341}%
  \BibitemOpen
  \bibfield  {author} {\bibinfo {author} {\bibfnamefont {B.~W.}\ \bibnamefont
  {Drinkwater}}\ and\ \bibinfo {author} {\bibfnamefont {P.~D.}\ \bibnamefont
  {Wilcox}},\ }\href {\doibase DOI 10.1016/j.ndteint.2006.03.006} {\bibfield
  {journal} {\bibinfo  {journal} {Ndt I\& E International}\ }\textbf {\bibinfo
  {volume} {39}},\ \bibinfo {pages} {525} (\bibinfo {year} {2006})}\BibitemShut
  {NoStop}%
\bibitem [{\citenamefont {Giurgiutiu}(2005)}]{RN3471}%
  \BibitemOpen
  \bibfield  {author} {\bibinfo {author} {\bibfnamefont {V.}~\bibnamefont
  {Giurgiutiu}},\ }\href {\doibase 10.1177/1045389x05050106} {\bibfield
  {journal} {\bibinfo  {journal} {Journal of Intelligent Material Systems I\&
  Structures}\ }\textbf {\bibinfo {volume} {16}},\ \bibinfo {pages} {291}
  (\bibinfo {year} {2005})}\BibitemShut {NoStop}%
\bibitem [{\citenamefont {Raghavan}\ and\ \citenamefont
  {Cesnik}(2007)}]{RN3596}%
  \BibitemOpen
  \bibfield  {author} {\bibinfo {author} {\bibfnamefont {A.~C.}\ \bibnamefont
  {Raghavan}}\ and\ \bibinfo {author} {\bibfnamefont {C.~E.~S.}\ \bibnamefont
  {Cesnik}},\ }\href@noop {} {\bibfield  {journal} {\bibinfo  {journal} {Shock
  I\& Vibration Digest}\ }\textbf {\bibinfo {volume} {39}},\ \bibinfo {pages}
  {91} (\bibinfo {year} {2007})}\BibitemShut {NoStop}%
\bibitem [{\citenamefont {Wilcox}\ \emph {et~al.}(2007)\citenamefont {Wilcox},
  \citenamefont {Konstantinidis}, \citenamefont {Croxford},\ and\ \citenamefont
  {Drinkwater}}]{RN3603}%
  \BibitemOpen
  \bibfield  {author} {\bibinfo {author} {\bibfnamefont {P.~D.}\ \bibnamefont
  {Wilcox}}, \bibinfo {author} {\bibfnamefont {G.}~\bibnamefont
  {Konstantinidis}}, \bibinfo {author} {\bibfnamefont {A.~J.}\ \bibnamefont
  {Croxford}}, \ and\ \bibinfo {author} {\bibfnamefont {B.~W.}\ \bibnamefont
  {Drinkwater}},\ }\href@noop {} {\bibfield  {journal} {\bibinfo  {journal}
  {Proceedings Mathematical Physical I\& Engineering Sciences}\ }\textbf
  {\bibinfo {volume} {463}},\ \bibinfo {pages} {2961} (\bibinfo {year}
  {2007})}\BibitemShut {NoStop}%
\bibitem [{\citenamefont {Kessler}\ \emph {et~al.}(2002)\citenamefont
  {Kessler}, \citenamefont {Spearing},\ and\ \citenamefont {Soutis}}]{RN3595}%
  \BibitemOpen
  \bibfield  {author} {\bibinfo {author} {\bibfnamefont {S.~S.}\ \bibnamefont
  {Kessler}}, \bibinfo {author} {\bibfnamefont {S.~M.}\ \bibnamefont
  {Spearing}}, \ and\ \bibinfo {author} {\bibfnamefont {C.}~\bibnamefont
  {Soutis}},\ }\href {\doibase Pii S0964-1726(02)33619-X Doi
  10.1088/0964-1726/11/2/310} {\bibfield  {journal} {\bibinfo  {journal} {Smart
  Materials I\& Structures}\ }\textbf {\bibinfo {volume} {11}},\ \bibinfo
  {pages} {269} (\bibinfo {year} {2002})}\BibitemShut {NoStop}%
\bibitem [{\citenamefont {Redwood}(1958)}]{RN3163}%
  \BibitemOpen
  \bibfield  {author} {\bibinfo {author} {\bibfnamefont {M.}~\bibnamefont
  {Redwood}},\ }\href {\doibase Doi 10.1088/0370-1328/72/5/320} {\bibfield
  {journal} {\bibinfo  {journal} {Proceedings of the Physical Society of
  London}\ }\textbf {\bibinfo {volume} {72}},\ \bibinfo {pages} {841} (\bibinfo
  {year} {1958})}\BibitemShut {NoStop}%
\bibitem [{\citenamefont {Redwood}(1959)}]{RN3167}%
  \BibitemOpen
  \bibfield  {author} {\bibinfo {author} {\bibfnamefont {M.}~\bibnamefont
  {Redwood}},\ }\href {\doibase Doi 10.1121/1.1907732} {\bibfield  {journal}
  {\bibinfo  {journal} {Journal of the Acoustical Society of America}\ }\textbf
  {\bibinfo {volume} {31}},\ \bibinfo {pages} {442} (\bibinfo {year}
  {1959})}\BibitemShut {NoStop}%
\bibitem [{\citenamefont {Redwood}(1960)}]{RN3400}%
  \BibitemOpen
  \bibfield  {author} {\bibinfo {author} {\bibfnamefont {M.}~\bibnamefont
  {Redwood}},\ }\href@noop {} {\emph {\bibinfo {title} {Mechanical waveguides :
  the propagation of acoustic and ultrasonic waves in fluids and solids with
  boundaries}}}\ (\bibinfo  {publisher} {Pergamon Press},\ \bibinfo {year}
  {1960})\BibitemShut {NoStop}%
\bibitem [{\citenamefont {Greve}\ \emph {et~al.}(2007)\citenamefont {Greve},
  \citenamefont {Zheng},\ and\ \citenamefont {Oppenheim}}]{RN3479}%
  \BibitemOpen
  \bibfield  {author} {\bibinfo {author} {\bibfnamefont {D.~W.}\ \bibnamefont
  {Greve}}, \bibinfo {author} {\bibfnamefont {P.}~\bibnamefont {Zheng}}, \ and\
  \bibinfo {author} {\bibfnamefont {I.~J.}\ \bibnamefont {Oppenheim}},\ }\href
  {\doibase Doi 10.1109/Ultsym.2007.283} {\bibfield  {journal} {\bibinfo
  {journal} {2007 Ieee Ultrasonics Symposium Proceedings, Vols 1-6}\ ,\
  \bibinfo {pages} {1120}} (\bibinfo {year} {2007})}\BibitemShut {NoStop}%
\bibitem [{\citenamefont {Greve}\ \emph {et~al.}(2008)\citenamefont {Greve},
  \citenamefont {Zheng},\ and\ \citenamefont {Oppenheim}}]{RN1462}%
  \BibitemOpen
  \bibfield  {author} {\bibinfo {author} {\bibfnamefont {D.~W.}\ \bibnamefont
  {Greve}}, \bibinfo {author} {\bibfnamefont {P.}~\bibnamefont {Zheng}}, \ and\
  \bibinfo {author} {\bibfnamefont {I.~J.}\ \bibnamefont {Oppenheim}},\ }\href
  {\doibase Artn 035029 10.1088/0964-1726/17/3/035029} {\bibfield  {journal}
  {\bibinfo  {journal} {Smart Materials I\& Structures}\ }\textbf {\bibinfo
  {volume} {17}} (\bibinfo {year} {2008}),\ Artn 035029
  10.1088/0964-1726/17/3/035029}\BibitemShut {NoStop}%
\bibitem [{\citenamefont {Rose}(2014)}]{Rose2014Ultrasonic}%
  \BibitemOpen
  \bibfield  {author} {\bibinfo {author} {\bibfnamefont {J.~L.}\ \bibnamefont
  {Rose}},\ }\href@noop {} {\emph {\bibinfo {title} {Ultrasonic guided waves in
  solid media}}}\ (\bibinfo  {publisher} {Cambridge University Press},\
  \bibinfo {year} {2014})\BibitemShut {NoStop}%
\end{thebibliography}%

\end{document}